\documentclass[prd,twocolumn,superscriptaddress,floatfix,showpacs,10pt,letterpaper]{revtex4}
\usepackage{graphicx} \usepackage{subfigure} \usepackage{amssymb}
\usepackage{verbatim} \usepackage{amsmath} \usepackage{amscd}
\usepackage{amssymb} 
\usepackage{epsfig}     

\newcommand{\p}{x_+}
\newcommand{\m}{x_-}
\newcommand{\g}{g_{+-}}

\newcommand{\rd}[1]{\mathop{\mathrm{d}#1}}
\newcommand{\diff}{\mathcal{L}_{\xi}}
\newcommand{\met}{g_{\mu\nu}}
\newcommand{\exo}{\theta_{(l)}}
\newcommand{\ar}{a_{\Delta}}
\newcommand{\AdS}{\mathrm{AdS}_3}

\begin{document}

\title{Dynamics of Diffeomorphism Degrees of Freedom at a Horizon}

\author{Hyeyoun Chung}  
\email{hyeyoun@physics.harvard.edu.}
\affiliation{Jefferson Physical Laboratory, Harvard University,\\ 17 Oxford St., Cambridge, MA 02138, USA}

\date{\today} \begin{abstract} We define a set of boundary conditions that ensure the presence of a null hypersurface with the essential characteristics of a horizon, using the formalism of weakly isolated horizons as a guide. We then determine the diffeomorphisms that preserve these boundary conditions, and derive a dynamical action for these diffeomorphisms in a neighbourhood of the horizon. The action is similar to that of Liouville theory, and the equation of motion of the gravitational degrees of freedom approaches that of a free two-dimensional conformal field in the near-horizon region.\end{abstract}

 \maketitle 

\section{Introduction} 

Ever since the work of Bekenstein\cite{Bekenstein}, who posited that black hole horizons have entropy, and Hawking\cite{Hawking}, who showed that horizons can emit radiation, it has generally been accepted that black holes are thermodynamic objects. It was later shown that these characteristics are shared by other types of horizons, such as cosmological deSitter horizons and even acceleration horizons\cite{Unruh}.

A central problem in studying the thermodynamic properties of horizons has been to determine the microscopic theory that accounts for the entropy and temperature of horizons, as well as the phenomenon of Hawking radiation. The derivation of the Bekenstein-Hawking entropy is generally taken to be a criterion for testing the validity of a theory of quantum gravity, and in recent years the well-known formula for the entropy of a horizon has been obtained using a wide variety of techniques (see \cite{CarlipReview} for a review). These techniques are based on many different theories, such as string theory\cite{StromingerVafa}, and loop quantum gravity\cite{LQG}, but none of them is wholly satisfactory. For example, computations using loop quantum gravity depend on a parameter called the Barbero-Immirzi parameter, whose value must be given as an additional input into the theory. String theory computations have the disadvantage that they must be carried out separately for each different type of black hole: successfully computing the entropy of a black hole in five dimensions does not necessarily give any information about the entropy of a black hole in four dimensions. 

The very fact that so many approaches to the problem of calculating the horizon entropy converge to the same answer, raises the possibility that there is an effective description of the gravitational degrees of freedom at a horizon that does not depend on the details of quantum gravity. The work of Strominger and Carlip, among others, in using the Cardy formula to calculate horizon entropy is strong evidence in favour of this idea\cite{StromingerCarlip}. The applicablity of the Cardy formula relies on determining the symmetries that govern the gravitational degrees of freedom in a spacetime with a horizon\cite{BrownHenneaux}. Most works that have used this approach to derive the Bekenstein-Hawking entropy have used physically motivated boundary conditions at a horizon or at infinity to find these symmetries, by deriving the diffeomorphisms that preserve these boundary conditions\cite{Dreyer, Silva, Koga, Kang}. The striking feature of this method is that a \textit{classical} conformal symmetry can be enough to determine the entropy, without knowing anything about the underlying quantum theory, lending support to the notion that there is an effective description of the horizon degrees of freedom that is independent of the true theory of quantum gravity. Moreover, the appearance of the Virasoro algebra suggests that this effective theory is a two-dimensional conformal field theory (CFT).

There have also been attempts to determine the precise nature of the degrees of freedom at a horizon. It has been suggested by Carlip that these degrees of freedom are ``would-be gauge'' degrees of freedom: that is, diffeomorphisms that become dynamical at the horizon due to the presence of boundaries or constraints. Dynamical actions have been derived for such would-be gauge degrees of freedom at spatial infinity in $\AdS$ and $\mathrm{AdS}_5$, with the former yielding a Liouville field theory\cite{CarlipAdS}, and the latter a theory of four-dimensional conformal gravity\cite{Aros}. Carlip has also derived the horizon entropy of the (2+1)-dimensional BTZ black hole by deriving an action for these ``would-be gauge'' degrees of freedom at the black hole horizon\cite{Carlip21}.

In this work we identify similar degrees of freedom that arise at a general horizon, by isolating diffeomorphisms that preserve the characteristics of the horizon, and finding a dynamical action for these diffeomorphisms in the near-horizon limit. Most work on horizon entropy has concentrated on black hole horizons, but there is good reason to believe that all horizons have thermodynamic properties\cite{Padmanabhan}. We therefore use a formalism that covers a very wide class of horizons. We find that the dynamical action for the gravitational degrees of freedom describes a two-dimensional theory that becomes a free 2D CFT in an infinitesimal neighbourhood of the horizon. The action is similar in form to the Liouville action. 

Earlier works have found a Liouville action for the gravitational degrees of freedom at a horizon, either choosing the dilaton field\cite{Solodukhin, DiasLemos}, or the conformal factor\cite{Giacomini} to be the Liouville field. Both derivations assume that the metric is spherically symmetric, and begin with an \textit{ad hoc} dimensional reduction to the $r-t$ plane. Furthermore, in \cite{Giacomini} the final action is not obtained from the Einstein-Hilbert action, but rather chosen so that it will lead to the equation of motion for the conformal field. Another approach has been to model the near-horizon region of black holes with a 2-dimensional Liouville-like effective theory, which is used to derive the Bekenstein-Hawking entropy and the total flux of Hawking radiation\cite{Rodriguez}. In contrast to these works, this paper directly relates the gravitational degrees of freedom at the horizon to the diffeomorphisms that preserve horizon boundary conditions, and we only integrate over the spatial coordinates as a final step, after determining that the leading order dynamics are in the $r-t$ plane. Lastly, we do not require the metrics to be spherically symmetric. 

This paper is structured as follows. In Sec. \ref{sec-IH} we review the formalism of isolated horizons, which will be useful in defining boundary conditions at a horizon. In Sec. \ref{sec-GN} we describe the coordinate systems that we will use in this paper. In Sec. \ref{sec-BC} we formulate the boundary conditions that preserve the existence and characteristics of a horizon, and provide physical motivations for these boundary conditions. In Sec. \ref{sec-Action} we derive the dynamical action for the diffeomorphism degrees of freedom at a horizon that preserve these boundary conditions, given certain conditions on the energy-momentum tensor of any matter fields present. In Sec. \ref{sec-Bound} we confirm that our boundary conditions lead to a well-defined action principle. In Sec. \ref{sec-General} we generalize our results to include a larger class of horizons. We conclude in Sec. \ref{sec-Conclusion}.

\section{Isolated Horizons}\label{sec-IH}

In this section we review the concept of isolated horizons, which will be useful in setting up boundary conditions that determine the existence and properties of a horizon. As we are concerned with the local degrees of freedom in a neighbourhood of the horizon, we will use the definition of \textbf{weakly isolated horizons} as a guide. This formalism describes horizons that are in equilibrium, and has several features that make it suitable for our purposes. We largely follow the approach of \cite{Ashtekar} and \cite{Booth}.

Isolated horizons are null sub-manifolds $\Delta$ of spacetime, with an intrinsic metric $q_{ab}$ that is the pull-back of the spacetime metric to $\Delta$. A tensor $q^{ab}$ on $\Delta$ is defined to be an inverse of $q_{ab}$ if it satisfies $q_{am}q_{bn}q^{mn}=q_{ab}$. The inverse is not unique, but all of the definitions and constructions in the isolated horizon formalism are independent of the choice of inverse. Given a null normal $l^\mu$ to $\Delta$, the \textit{expansion} $\theta_{(l)}$ of $l^\mu$ is defined to be
\begin{equation}
\theta_{(l)} := q^{ab}\nabla_a l_b.
\end{equation}
A weakly isolated horizon (WIH) is a sub-manifold $\Delta$ of a spacetime that satisfies the following conditions:
\begin{enumerate}
\item $\Delta$ is topologically $S^2\times\mathbb{R}$ and null,

\item Any null normal $l^\mu$ of $\Delta$ has vanishing expansion, $\theta_{(l)}=0$, and

\item All equations of motion hold at $\Delta$ and the stress energy tensor $T_{\mu\nu}$ is such that $-T^\mu_\nu l^\nu$ is future-causal for any future directed null normal $l^\mu$.
\end{enumerate}
Note that if Condition 2 holds for one null normal to $\Delta$, then it holds for all. WIHs generalize the definitions of Killing horizons and apparent horizons, with the normal vector $l^\mu$ being analogous to a Killing vector, and the requirement that $\exo=0$ clearly being inspired by the notion of trapped surfaces. However, the definition of a WIH is given only \textit{at} the horizon, and does not require a Killing vector to exist even within an infinitesimal neighborhood. Thus, WIHs allow for much greater freedom in the dynamics of the matter and spacetime outside the horizon, while preserving the essential characteristics of the horizon itself.

The surface gravity of an isolated horizon is not uniquely defined. However, given a normal vector $l^\mu$ to $\Delta$, there is a function $\kappa_l$ such that
\begin{equation}\label{eq-sgra}
l^\mu\nabla_\mu l^\nu = \kappa_l l^\nu.
\end{equation}
It is always possible to choose a normal vector $l^\mu$ such that the corresponding $\kappa_l$ is constant everywhere on $\Delta$. Therefore, $\kappa_l$ may be interpreted as the surface gravity of the horizon corresponding to $l^\mu$.

\section{Gaussian Null Coordinates and Conformal Coordinates.}\label{sec-GN}

In this section we introduce the coordinate systems that we use in this work. The first set of coordinates are known as \textbf{Gaussian null coordinates} (denoted ``GN coordinates''), and are analogous to Eddington-Finkelstein coordinates in Schwarzschild spacetime\cite{Wald2}. In the neighbourhood of any smooth null hypersurface $\Delta$, it is possible to define GN coordinates $(u,r,x^i)$ such that the metric takes the form
\begin{equation}\label{eq-GN}
{\rd{s}}^2 = rF{\rd{u}}^2 + 2\rd{u}\rd{r}+2rh_i\rd{u}\rd{x^i}+g_{ij}\rd{x^i}\rd{x^j},
\end{equation}
where $g_{ij}$ is positive definite, and $F, h_i,$ and $g_{ij}$ are smooth functions of $(u,r,x^i)$ that can be expanded in powers of $r$. The null hypersurface is defined by $r=0$, and we have chosen a smooth, non-vanishing vector field $l^\mu$ that is normal to $\Delta$, so that the integral curves of $l^\mu$ are the null geodesic generators of $\Delta$ and we have $l^\mu= (\partial/\partial u)^\mu$ on $\Delta$.

Since an isolated horizon is a null hypersurface, we can construct such a coordinate system in a neighbourhood of any isolated horizon. In fact, in the neighbourhood of the event horizon of a stationary black hole, or a stationary Killing horizon, we can define these coordinates so that all the metric components are independent of $u$. Extremal Killing horizons correspond to the case where $F$ has a simple root at $r=0$, so that it has the form $F=rf(u,r,x^i)$. We will consider only \textit{non-extremal} horizons, such that $F|_{r=0}\neq 0$. We can then define a non-zero surface gravity for the horizon, using the definition (\ref{eq-sgra}) and taking the normal vector to the $r=0$ hypersurface to be $l^\mu := g^{r\mu}$. The surface gravity $\kappa$ associated with this normal vector is $-\frac{1}{2}F|_{r=0}$.

To illustrate our general method, from now on we restrict ourselves to cases where $h_i=0$, and $F=F(u,r)$, so that the metric takes the form
\begin{equation}\label{eq-GN2}
{\rd{s}}^2 = rF{\rd{u}}^2 + 2\rd{u}\rd{r}+g_{ij}\rd{x^i}\rd{x^j}.
\end{equation}
Later we will return to study a much more general case, where $h_i \neq 0$ and $F$ and $h_i$ are functions of $(u,r,x^i)$. We will find that our results can be easily generalized, so the current simplification is made only for clarity of presentation: it will enable us to explain the important principles of our work while ignoring inessential details. Considering the simpler class of metrics of the form (\ref{eq-GN2}), we can define conformal coordinates $(\p,\m)$ in terms of the coordinates $(u,r)$ such that the metric takes the form
\begin{equation}
{\rd{s}}^2 = 2\g\rd{\p}\rd{\m}+g_{ij}\rd{x^i}\rd{x^j},\label{eq-conf}
\end{equation}
where $g_{+-}=e^{\sigma(u,r)}$ with $\sigma(u,r) = \ln r+\sigma_0(u)+O(r)$. Without loss of generality, we can require the GN coordinates and the conformal coordinates to satisfy:
\begin{align}
\partial_- u &= 0,\quad\partial_r \p = 0.
\end{align}
We can also determine the useful relations:
\begin{align}
\partial_+r &= O(r),\quad\partial_-r = O(r).\label{eq-r}
\end{align}

\section{Horizon boundary conditions}\label{sec-BC}

In this section we use the definition of WIHs given in Sec. \ref{sec-IH} to formulate boundary conditions that preserve the existence and characteristics of a horizon, and express these boundary conditions in the GN coordinates and conformal coordinates described in Sec. \ref{sec-GN}. We then identify the diffeomorphisms that preserve these boundary conditions. These diffeomorphisms can be interpreted as the gravitational degrees of freedom at a horizon, and in Sec. \ref{sec-Action} we will derive a dynamical action for them. We will eventually find that the action is similar to the of Liouville theory, and the equation of motion of the gravitational degrees of freedom approaches that of a free two-dimensional conformal field in the near-horizon region.

To define the boundary conditions, we assume that we have a weakly isolated horizon $\Delta$ in our spacetime, and that we can define GN coordinates in a neighbourhood of $\Delta$ so that the horizon lies at $r=0$ and the metric takes the form in Eq.(\ref{eq-GN2}). The requirement that the horizon satisfy $\exo=0$ is equivalent to requiring $\partial_u g_{ij} = O(r)$. We now apply an infinitesimal diffeomorphism $\xi$, and obtain a new metric $g_{\mu\nu}' := g_{\mu\nu} + \diff\met$. We then impose conditions that preserve the essential characteristics of the horizon by demanding that, after the diffeomorphism:
\begin{enumerate}
\item There is still a null hypersurface at $r=0$. This is equivalent to saying that $g_{uu}'$ (or, in conformal coordinates, $\g'$) has a simple root at $r=0$. \label{cond-1}

\item In conformal coordinates, the metric remains in the form given by Eq.(\ref{eq-conf}), with $\g' = e^{\sigma'(u,r)}$ for some $\sigma'(u,r) = \ln r + \sigma_0'(u) + O(r)$.\label{cond-2}

\item The induced metric on the $r=0$ hypersurface is be preserved, so that $g'_{ij} = g_{ij} + O(r)$.\label{cond-3}
\end{enumerate}

Each of these boundary conditions is physically motivated. Condition \ref{cond-1} is necessary (but not sufficient) for a WIH to still exist at $r=0$ following the diffeomorphism. Condition \ref{cond-2} reflects the fact that the class of horizons we are currently considering can all be written in the form (\ref{eq-conf}): thus, if the horizon remains intact after the diffeomorphism then the metric should remain in the same form up to trivial diffeomorphisms. This condition is analogous to the boundary conditions defining asymptotically AdS spaces. The metric of any asymptotically AdS space can be written in a special form, called the Fefferman-Graham form\cite{FeffermanGraham}. Therefore the diffeomorphisms that keep the metric in Fefferman-Graham form are the dynamical degrees of freedom at the spatial infinity of an asymptotically AdS space\cite{CarlipAdS, Aros}. Finally, Condition \ref{cond-3} is derived from the requirement that $\exo=0$ for WIHs. In GN coordinates this requirement is equivalent to $\partial_u g_{ij}=O(r)$. It follows that the induced metric on the $r=0$ hypersurface is a characteristic of the horizon, as it remains constant along $\Delta$. We therefore impose that this characteristic of the horizon is preserved under the diffeomorphism $\xi$, so that $g_{ij}' = g_{ij} + O(r)$. This conditions also ensures that the null hypersurface at $r=0$ continues to satisfy $\exo=0$, so that we still have a WIH at $r=0$ following the diffeomorphism.

Working in conformal coordinates, we determine the diffeomorphisms that satisfy the above conditions, and find that they have the form:
\begin{align}\label{eq-diff}
\xi^+ &= \xi^+(\p) + O(r)\\
\xi^- &= \xi^-(\m) + O(r)\nonumber\\
\xi^i &= O(r)\nonumber
\end{align}
This form of $\xi$ is extremely suggestive: the $(\p,\m)$ coordinates define a natural two-dimensional submanifold where our CFT will live, with infinitesimal conformal transformations being given by $\p \to\p+\xi^+(\p)$, $\m \to\m+\xi^+(\m)$. We also see that the coordinate $\p + \m$ transforms like the Liouville field. In the case of a simple static horizon, this is the radial tortoise coordinate. Thus the radial degree of freedom may play the part of a Liouville-like field in the CFT at the horizon. Under this diffeomorphism, the metric transforms as:
\begin{align}\label{eq-newMet}
g'_{+-} &= g_{+-}(1+\partial_+\xi^+ +\partial_-\xi^-+\xi^+\partial_+\sigma+\xi^-\partial_-\sigma)\\
g'_{ij} &= g_{ij} + \xi^+\partial_+g_{ij} +\xi^-\partial_-g_{ij}\nonumber
\end{align}

\section{The Dynamical Action}\label{sec-Action}

In this section we derive a dynamical action for the gravitational degrees of freedom at the horizon. Our general procedure is as follows: we start with a background metric $g_{\mu\nu}$ in the form (\ref{eq-conf}). It is important to note that we do \textit{not} require $\met$ to satisfy Einstein's equations (EEs): we only require it to have a form such that it \textit{can} satisfy the EEs \textit{asymptotically} in a region infinitesimally close to the horizon, as $r\to 0$. We explain what this means more precisely below.

We will assume that the matter energy-momentum tensor $T_{\mu\nu}$ satisfies:
\begin{align}
T_{+-} \,\,\,&\hat{=}\,\,\, O(r^2)\label{eq-TAssump1}\\
T_{++} \,\,\,&\hat{=}\,\,\, O(r^2)\label{eq-TAssump2}\\
T_{--} \,\,\,&\hat{=}\,\,\, O(r^2)\label{eq-TAssump3}\\
T_{ij} \,\,\,&\hat{=}\,\,\, O(r),\label{eq-TAssump4}
\end{align}
where ``$\hat{=}$'' indicates that the constraint holds as $r\to0$ (We will return to these assumptions at the end of this section and consider if they are reasonable. This is certainly the case if there is no matter in the near-horizon region, even if there are arbitrary distributions of matter fields elsewhere.) Looking at the Einstein equation:
\begin{equation}\label{eq-EE}
R_{\mu\nu} - \frac{1}{2}Rg_{\mu\nu} + \Lambda g_{\mu\nu} = 8\pi G T_{\mu\nu}
\end{equation}
for metrics of the form (\ref{eq-conf}), this gives the following conditions on the Ricci tensor:
\begin{align}
R_{+-} \,\,\,&\hat{=}\,\,\, O(r)\label{eq-Ricci1}\\
R_{++} \,\,\,&\hat{=}\,\,\, O(r^2)\label{eq-Ricci2a}\\
R_{--} \,\,\,&\hat{=}\,\,\, O(r^2)\label{eq-Ricci2b}\\
R \,\,\,&\hat{=}\,\,\, \frac{2n\Lambda}{n-2} + O(r)\label{eq-Ricci3}
\end{align}
Evaluating the Ricci tensor for metrics of the form (\ref{eq-conf}), we find:
\begin{align}
R_{+-} &= -\partial_+\partial_-\sigma + O(r)\label{eq-Ricci4}\\
R_{++} &= -g^{ik}(\partial_+^2 g_{ik} + \partial_+\sigma\partial_+ g_{ik}) + O(r^2)\label{eq-Ricci5a}\\
R_{--} &= -g^{ik}(\partial_-^2 g_{ik} + \partial_-\sigma\partial_- g_{ik}) + O(r^2)\label{eq-Ricci5b}\\
R_{ij} &= \tilde{R}_{ij} - g^{+-}\partial_+\partial_-g_{ij} + O(r),\label{eq-Ricci6}
\end{align}
where $\tilde{R}_{ij}$ denotes the Ricci tensor derived from the induced metric $\tilde{g}_{ij} := g_{ij}|_{r=0}$ on the $r=0$ hypersurface. From now on, given any tensor $A_{\mu_1\dots\mu_k}^{\nu_1\dots\nu_l}$, the notation $\tilde{A}_{\mu_1\dots\mu_k}^{\nu_1\dots\nu_l}$ will denote the corresponding tensor computed using $\tilde{g}_{ij}$. Details of the calculation are given in Appendix \ref{app-1}.

Comparing Eq.(\ref{eq-Ricci1})-(\ref{eq-Ricci3}) to Eq. (\ref{eq-Ricci4})-(\ref{eq-Ricci6}) shows that in order for $\met$ to be capable of satisfying the EEs as $r\to 0$, we must have
\begin{align}
\partial_+\partial_-\sigma \,\,\,&\hat{=}\,\,\, O(r)\label{eq-constraints1}\\
-g^{ik}(\partial_+^2 g_{ik} + \partial_+\sigma\partial_+ g_{ik}) \,\,\,&\hat{=}\,\,\, O(r^2)\label{eq-constraints2}\\
-g^{ik}(\partial_-^2 g_{ik} + \partial_-\sigma\partial_- g_{ik}) \,\,\,&\hat{=}\,\,\, O(r^2)\label{eq-constraints3}\\
R -\frac{2n\Lambda}{n-2} \,\,\,&\hat{=}\,\,\, O(r),\label{eq-constraints4}
\end{align}
The constraints (\ref{eq-constraints1})-(\ref{eq-constraints4}) are necessary, but not sufficient, for $\met$ to satisfy the EEs as $r\to 0$. Therefore, we are not requiring $\met$ to satisfy the EEs even in an infinitesimal neighborhood of the horizon.

We apply a diffeomorphism $\xi$ of the form (\ref{eq-diff}) to the background metric $\met$, obtaining a new metric $\met' := \met + \diff\met$. We then evaluate the Einstein-Hilbert action for the new metric $\met'$ with the constraints (\ref{eq-constraints1})-(\ref{eq-constraints4}) applied to $\met$ in the near-horizon region, thus isolating the gravitational fluctuations about this background that preserve the horizon. When evaluating the Einstein-Hilbert action, everything is calculated from the new metric, including the inverse metric and the metric determinant, as the metric itself is the only dynamical field in the problem. The final form of the action determines the dynamics of the horizon degrees of freedom in an infinitesimal neighborhood of the horizon. As we do not require $\met$ to satisfy the Einstein equations, we get a non-trivial form for the action.

We begin with the Einstein-Hilbert action
\begin{equation}
I_{EH} = \frac{1}{16\pi G} \int\rd{^nx}  \sqrt{-g}\,\,(R-2\Lambda),
\end{equation}
and evaluate the action for the new metric $\met'$. We see from (\ref{eq-newMet}) that this metric has the form $\g' = (1+\phi)\g$ and $g_{ij}' = g_{ij} + O(r)$, with $\phi$ given by
\begin{equation}
\phi = \partial_+\xi^+ +\partial_-\xi^-+\xi^+\partial_+\sigma+\xi^-\partial_-\sigma.
\end{equation}
This $\phi$ will end up being the dynamical degree of freedom in the near-horizon region. As we are interested in the near-horizon region $r\to 0$, and we are considering infinitesimal diffeomorphisms, we work to leading order in $(r,\phi)$. To second order in $\phi$, the inverse metric ${g'}^{\mu\nu}$ has the form
\begin{align}
{g'}^{+-} &= (1-\phi + \phi^2)g^{+-}\\
{g'}^{ij} &= g^{ij} + O(r)
\end{align}
We can now begin evaluating each of the terms necessary to compute the Einstein-Hilbert action. First we find that
\begin{align}
\sqrt{-g'} &= \g'\sqrt{\tilde{g}} + O(r^2).
\end{align}
Evaluating the Ricci tensor gives:
\begin{align}
R'_{+-} &= -{g'}^{+-}\partial_+\partial_-\g' + ({g'}^{+-})^2\partial_-\g'\partial_+\g'\nonumber\\
&\,\,\,\,\,\,\,\,\,-\frac{1}{2}{g'}^{ij}\partial_-\partial_+ g'_{ij} + O(r^2)\label{eq-RicciNew}
\end{align}
We can express this in terms of the field $\phi$, and simplify the result by applying the constraints (\ref{eq-constraints1})-(\ref{eq-constraints4}) to the background metric $\met$. The details of the calculation are given in Appendix \ref{app-2}. We obtain:
\begin{align}\label{eq-EHTerm1}
R'_{+-} &= \partial_+\phi\partial_-\phi\nonumber\\
&\hspace{0.4cm}+ (1+\partial_-\xi^- + \partial_+\xi^+)\partial_+\partial_-\sigma\nonumber\\
&\hspace{0.4cm}+ \xi^+\partial_+^2\partial_-\sigma + \xi^-\partial_+\partial_-^2\sigma\nonumber\\
&\hspace{0.4cm}-\frac{1}{2}{g'}^{ij}\partial_-\partial_+ g'_{ij} + O(r^2)+ O(\phi^3, r\phi^2)
\end{align}
We can also compute
\begin{align}
R_{ij}' &= \tilde{R}'_{ij} - {g'}^{+-}\partial_+\partial_-g'_{ij} + O(r)
\end{align}
which can be simplified using results in Appendix \ref{app-2} to give
\begin{align}\label{eq-EHTerm2}
{g'}^{ij}R'_{ij} &= g^{ij}\tilde{R}_{ij}- {g'}^{ij}{g'}^{+-}\partial_+\partial_-g'_{ij}+ O(r)
\end{align}
Putting together Eq. (\ref{eq-EHTerm1}) and (\ref{eq-EHTerm2}) to compute the entire Einstein-Hilbert action, we get:
\begin{align}\label{eq-newR}
\sqrt{-g'} (R' - 2\Lambda) &= \sqrt{-g'}\left [ 2{g'}^{+-}R'_{+-} + {g'}^{ij}R'_{ij} \right ]\nonumber\\
&= \sqrt{\tilde{g}}\sqrt{-\hat{g}}\left [ \hat{g}^{ab}\partial_a\phi\partial_b\phi -\phi\hat{R} + \lambda(1+\phi)\right ]\nonumber\\
&= \sqrt{\tilde{g}}\sqrt{-\hat{g}}\left [ \partial_a\phi\partial^a\phi -\phi\hat{R} + \lambda(1+\phi)\right ]
\end{align}
where the index $a \in (+,-)$, $\hat{g}$ is the induced metric on the $(\p,\m)$ submanifold, and $\lambda :=\frac{4\Lambda}{n-2}$. (In the case $n=2$ we have $\Lambda=0$.) The Ricci scalar $\hat{R}$ is computed from $\hat{g}$. Once again, details of the calculation are given in Appendix \ref{app-2}.

The first term in (\ref{eq-newR}) is a kinetic term for $\phi$, of order $O(\phi^2)$. The second term is a coupling term between $\phi$ and the induced Ricci scalar $\hat{R}$ on the $(\p,\m)$ submanifold. As this term is multiplied by $\sqrt{-\hat{g}}=O(r)$, it is of order $O(r\phi)$. Finally, we have a potential term of the form $(1+\phi)$, which is also multiplied by $\sqrt{-\hat{g}}$ and is thus of order $O(r)$. All the remaining terms in (\ref{eq-newR}) are higher order in $(r,\phi)$ and are therefore not shown, as they are sub-leading as $r\to 0$. We find that all dependence on the coordinates $x^i$ disappears except for the metric determinant $\sqrt{\tilde{g}}$, so we can simply integrate over the $x^i$ coordinates when calculating the action. We finally obtain
\begin{equation}\label{eq-action}
I_{\mathrm{hor}} = \frac{\ar}{16\pi G} \int\rd{^2x} \sqrt{-\hat{g}}\,\, \left (\partial_a\phi\partial^a\phi -\phi\hat{R}+ \lambda (1+\phi)\right )
\end{equation}
where $\ar$ is the cross-sectional area of the horizon, and the variables of integration are $(\p,\m)$. We note that the action $I_{\mathrm{hor}}$ is very similar to the familiar Liouville action with a background metric $\hat{g}_{ab}$. The only differences are the coefficient of the $\phi\hat{R}$ coupling, and the fact that we have a term of the form $(1+\phi)$ instead of the familiar $e^\phi$ term in the Liouville action, as we are considering infinitesimal diffeomorphisms. 

The dynamics described by $I_{\mathrm{hor}}$ are very simple. The non-kinetic terms in the action, although we have included them in the Lagrangian, are $O(r)$ and therefore become irrelevant in the near-horizon region. The equation of motion for $\phi$ has the form
\begin{equation}
\partial_+\partial_-\phi +O(r) = 0,
\end{equation}
so that $\phi$ becomes a free two-dimensional conformal field in an infinitesimal neighbourhood of the horizon. 

This is our main result: we have isolated the diffeomorphism degrees of freedom in a neighbourhood of a horizon that preserve the essential characteristics of the horizon, and derived a dynamical action for these degrees of freedom. We find that the gravitational fluctuations in the neighbourhood of a horizon are described by a two-dimensional field theory similar to Liouville theory that becomes conformal in the near-horizon region.

Our result clearly holds in all cases where there is no matter in the near-horizon region. We can now consider whether the assumptions (\ref{eq-TAssump1})-(\ref{eq-TAssump4}) are justified even when the near-horizon region contains matter. Recall that we are starting with a background field configuration defined by the metric $\met$ and any matter fields $\psi_m$ that are present in our spacetime, and $T_{\mu\nu}$ is evaluated on the background values of these fields. We assume that all the fields can be expanded in powers of $r$, so that a general field can be written as $\psi_m = \psi_m^0(x^i)+r\psi_m^1(x^i)+O(r^2)$ in the near-horizon region.

We first note that for metrics of the form (\ref{eq-conf}), assumptions (\ref{eq-TAssump2})-(\ref{eq-TAssump3}) are certainly very reasonable. A typical term in the expression for $T_{++}$ or $T_{--}$ will either be proportional to $g_{++}$ or $g_{--}$ and thus vanish, or contain two derivatives of the matter fields with respect to $x_+$ or $x_-$, which will give a term of $O(r^2)$ according to (\ref{eq-r}). Further evidence for the reasonableness of this constraint is provided by the work of Medved et al.\cite{Medved1, Medved2}, who have shown that purely geometric constraints in the near-horizon region of a generic stationary horizon allow $T_{\mu\nu}$ to be written in block-diagonal form such that $T_{\mu\nu} \propto g_{\mu\nu}$ for $\mu,\nu \in \{+,-\}$ and $T_{+i}=T_{-i} = 0$. Thus for a generic stationary horizon we will actually have $T_{++}=T_{--} =0$.

Assumptions (\ref{eq-TAssump1}) and (\ref{eq-TAssump4}) are more restrictive. Looking at the form of the metric in (\ref{eq-conf}), and the Einstein equations (\ref{eq-EE}), we see that we are basically requiring the matter perturbations to be small enough in the near-horizon region so that they have a sub-leading effect on the dynamics of the metric. If we no longer impose these assumptions, then the Einstein equations give
\begin{equation}
T_{+-} = O(g_{+-}) \,\,\,\hat{=}\,\,\, O(r),
\end{equation}
and in general we expect $T_{ij} \,\hat{=}\, O(1)$. Thus there may be background field configurations that do not satisfy our initial assumptions. But even in this more general case, we see that we still preserve the conditions (\ref{eq-Ricci1})-(\ref{eq-Ricci2b}) on the Ricci tensor, and therefore we obtain the same constraints (\ref{eq-constraints1})-(\ref{eq-constraints3}) on the metric as before. The only constraint that changes is (\ref{eq-constraints4}), which becomes:
\begin{align}
R - \frac{2n\Lambda-16\pi G T^\mu_\mu}{n-2} \,\,\,\hat{=}\,\,\, O(r)
\end{align}
except in the case $n=2$, when we have $R\,\hat{=}\,O(r)$ as before. Working through the derivation of the dynamical action, we find that only the constraints (\ref{eq-constraints1})-(\ref{eq-constraints3}) are required until the very end of the computation, and thus we obtain almost the same result as before:
\begin{align}
\sqrt{-g'} (R' - 2\Lambda) &= \sqrt{\tilde{g}}\sqrt{-\hat{g}}\biggl [ \partial_a\phi\partial^a\phi -\phi\hat{R} + \lambda(1+\phi)\nonumber\\
&\hspace{2cm}- \frac{16\pi G}{n-2}T^{\mu}_\mu(1+\phi)\biggr ]
\end{align}
Comparing this to (\ref{eq-newR}), we see that we once again obtain a Liouville-like action, but with a modified potential term for $\phi$. We would have to know the precise form of $T_{\mu\nu}$ before proceeding further. It is important to note, however, that the dynamics of the field $\phi$ are unchanged by the addition of matter. The non-kinetic terms in the action are still $O(r)$, and therefore become irrelevant in the near-horizon region, so that $\phi$ is still a free two-dimensional conformal field in an infinitesimal neighborhood of the horizon.

\section{Boundary terms in the action}\label{sec-Bound}

In this section we confirm that the boundary conditions defined in Sec. \ref{sec-BC} allow us to define a gravitational action with a well-defined variational principle. The boundary conditions at $r=0$ are different from the usual boundary conditions imposed at spatial infinity, and therefore the Einstein-Hilbert action may gain a new boundary term at $\Delta$ that is different from the usual Gibbons-Hawking term. It is known that the WIH boundary conditions lead to a well-defined action principle\cite{BoothVar, AshtekarVar}: however, we will derive the result independently, for two main reasons. Firstly, our boundary conditions are slightly different from the standard WIH boundary conditions, and thus may lead to a slightly different boundary term. And secondly, as the WIH formalism is usually presented using the spin-connection form of the metric, the results will be easier to understand if presented using our notation and coordinate system.

We can determine the boundary term by varying the Einstein-Hilbert action\cite{Wald1}. The horizon is a null hypersurface, but we can still apply Stokes's theorem as long as the normal vector $l^\alpha$ satisfies
\begin{equation}
\frac{1}{n}\epsilon_{\alpha_1 \dots \alpha_n} = l_{[\alpha_1}\tilde{\epsilon}_{\alpha_2\dots\alpha_n]}
\end{equation}
where $\tilde{\epsilon}$ is an induced volume form defined on $\Delta$ and $\epsilon$ is the natural volume form in the bulk.  Upon varying the Einstein-Hilbert action, we obtain the following term:
\begin{equation}
\int_{\mathcal{M}} \nabla_\alpha v^{\alpha} = \int_{\Delta} v_\alpha l^{\alpha},
\end{equation}
where we have
\begin{align}
v_\alpha l^\alpha &= l^\alpha g^{\beta\gamma} \left [\nabla_\gamma \left ( \delta g_{\alpha\beta}\right ) - \nabla_\alpha \left (\delta g_{\beta\gamma} \right ) \right ]
\end{align}
We can evaluate the boundary term in conformal coordinates for an arbitrary variation of the metric that is allowed under our horizon boundary conditions given in Sec. \ref{sec-BC}. We can choose $l^\alpha = (\partial/\partial u)^\alpha$, so that the only non-zero components of $l^\alpha$ are $l^+$ and $l^-$ in conformal coordinates. We obtain:
\begin{align}
v_\alpha l^{\alpha} &= l^+g^{+-}\left [\nabla_-(\delta g_{++}) - \nabla_+(\delta\g) \right ]\nonumber\\ 
&\,\,\,\,\,\,\,\,\,\,\,\,+ l^+g^{-+}\left [\nabla_+(\delta g_{+-}) - \nabla_+(\delta\g) \right ]\nonumber\\
&\,\,\,\,\,\,\,\,\,\,\,\,+ l^-g^{+-}\left [\nabla_-(\delta g_{+-}) - \nabla_-(\delta\g) \right ]\nonumber\\
&\,\,\,\,\,\,\,\,\,\,\,\,+ l^-g^{+-}\left [\nabla_+(\delta g_{--}) - \nabla_-(\delta\g) \right ]\nonumber\\
&\,\,\,\,\,\,\,\,\,\,\,\,+ l^+g^{ij}\left [\nabla_j(\delta g_{+i}) - \nabla_+(\delta g_{ij}) \right ]\nonumber\\
&\,\,\,\,\,\,\,\,\,\,\,\,+ l^-g^{ij}\left [\nabla_j(\delta g_{-i}) - \nabla_-(\delta g_{ij}) \right ]
\end{align}
Applying the horizon boundary conditions, we find that
\begin{align}
v_\alpha l^{\alpha} &= -l^+ g^{+-}\nabla_+(\delta \g) - l^-g^{+-}\nabla_-(\delta \g) + O(r)\nonumber
\end{align}
And so:
\begin{align}
\int_{\Delta} v_\alpha l^{\alpha} &= \int_{\Delta} -l^+ g^{+-}\nabla_+(\delta g_{+-}) -l^- g^{+-}\nabla_-(\delta g_{+-})\nonumber
\end{align}
We can calculate:
\begin{align}
\nabla_+(\delta\g) &= \partial_+(\delta\g) - \Gamma_{++}^\alpha(\delta g_{\alpha -}) - \Gamma_{+-}^\alpha(\delta g_{\alpha +})\nonumber\\
&= \partial_+(\delta\g) - \Gamma_{++}^+(\delta g_{+-})\nonumber\\
&= \g \partial_+(\partial_+\xi^+ + \partial_-\xi^- + \xi^+\partial_+\sigma + \xi^-\partial_-\sigma)\nonumber
\end{align}and similarly,
\begin{align}
\nabla_-(\delta\g) &= \g \partial_-(\partial_+\xi^+ + \partial_-\xi^- + \xi^+\partial_+\sigma + \xi^-\partial_-\sigma)\nonumber
\end{align}
This gives us:
\begin{align}
v_\alpha l^{\alpha} &= -l^+\partial_+(\partial_+\xi^+ + \dots) -l^-\partial_-(\partial_+\xi^+ + \dots)\nonumber\\
&= -\partial_u (\partial_+\xi^+ + \dots)
\end{align}
So finally we find:
\begin{align}
\int_{\Delta} v_\alpha l^{\alpha} &= \int_{\Delta}-\partial_u \left [g^{+-}\delta g_{+-} \right ]\nonumber\\
&= \int_{\Delta}-\frac{1}{2}\partial_u\delta\left [ \mathrm{Tr}(\hat{g})\right ]\nonumber\\
&= \delta \left[\frac{1}{2}\left(\mathrm{Tr}(\hat{g})|_{u_1} - \mathrm{Tr}(\hat{g})|_{u_2}\right) \right ]
\end{align}
where $\hat{g}$ is the induced metric on the $(\p,\m)$ sub-manifold, and $u_1$ and $u_2$ are the values of the $u$-coordinate at the endpoints of the segment $\Delta$. However, as the trace $\mathrm{Tr}(\hat{g})$ of $\hat{g}$ is constant along $\Delta$, the boundary term can actually be discarded altogether.

\section{More General Horizons}\label{sec-General}

Let us review what we have done so far. We have considered horizons that can be written in the form (\ref{eq-conf}), and isolated the diffeomorphisms that preserve the existence and essential characteristics of such a horizon. We have derived a dynamical action for these gravitational degrees of freedom and found that they are described by a Liouville-like theory in a neighborhood of the horizon, which becomes a free 2D CFT as the radial distance $r$ to the horizon approaches zero. 

In this section we show that a similar result holds for a much larger class of horizons, by considering the more general case where the GN coordinates in the near-horizon region yield a metric of the form (\ref{eq-GN}), with $h_i =h_i(u,r,x^i) \neq 0$, and $F = F(u,r,x^i)$. Once again we consider non-extremal horizons, with $F|_{r=0}\neq 0$.

As before, we can define coordinates $(\p,\m)$ in terms of the coordinates $(u,r)$ so that the metric takes the form
\begin{equation}\label{eq-conf2}
{\rd{s}}^2 = 2\g\left (\rd{\p}\rd{\m}+h_{+i}\rd{x^+}\rd{x^i}\right)+g_{ij}\rd{x^i}\rd{x^j},
\end{equation}
where $\g=e^{\sigma(u,r,x^i)}$, and $\sigma(u,r,x^i)=\ln r + \sigma_0(u,x^i) + O(r)$. The boundary conditions assuring the existence of a horizon at $r=0$ are almost the same as those given in Sec. \ref{sec-BC}, except that now the metric is required to remain in the form given by Eq.(\ref{eq-conf2}) rather than Eq.(\ref{eq-conf}). Also, in order to obtain a simple Liouville-type action, we have to impose one more restriction on the class of metrics under consideration: we require that $\partial_i\partial_+\sigma=O(r)$, so that $\sigma = \ln r + \sigma_0(u) + \sigma_1 (x^i) + O(r)$. As the form of the functions $h_{+i}$ are left unrestricted, this still allows us to describe a very large class of horizon metrics. For example, all stationary horizons satisfy these conditions.

We find that the diffeomorphisms preserving these boundary conditions have the same form as in Eq.(\ref{eq-diff}). This may seem surprising, as the form of the metric is now much less restricted. And indeed, almost all of the boundary conditions may be satisfied by more general diffeomorphisms of the form $\{ \xi^+(\p), \xi^-(\m,x^i), 0,0\}$. However, if the metric must remain in the form (\ref{eq-conf2}), then the factor $\g$ must be of the form $\g=e^{\ln r + \sigma_0(u,x^i) + O(r)}$, with $\partial_i\partial_-\sigma = O(r)$. The need to preserve this condition imposes $\partial_i\xi^- = O(r)$. These boundary conditions once again lead to a well-defined action principle, with a constant boundary term in the Einstein-Hilbert action that can be discarded as described in Sec. \ref{sec-Bound}. 

We derive the dynamical action for the horizon degrees of freedom in the same way as in Sec. \ref{sec-Action}, beginning with a background metric $\met$ of the form (\ref{eq-conf2}) and applying an infinitesimal diffeomorphism $\xi$ to obtain a new metric $\met' := \met + \mathcal{L}_{\xi}\met$. The new metric has the form
\begin{align}
\g' &= (1 + \partial_+\xi^+ + \partial_-\xi^- + \xi^+\partial_+\sigma + \xi^-\partial_-\sigma)\g\nonumber\\
g'_{+i} &= \partial_+\xi^+ g_{+i} + \partial_i\xi^-\g + \xi^-\partial_-g_{+i} + \xi^+\partial_+g_{+i}\nonumber\\
g'_{ij} &= \xi^+\partial_+g_{ij} + \xi^-\partial_-g_{ij}
\end{align}
As we have already outlined the general method in Sec. \ref{sec-Action}, we will give the details of the remaining calculations in Appendix \ref{app-3} and only give the main results here. Evaluating the Ricci tensor for metrics of the form (\ref{eq-conf2}), we find:
\begin{align}
R_{+-} &= -\partial_+\partial_-\sigma + O(r)\\
R_{-i} &= \frac{1}{2}(\partial_-^2\sigma h_{+i} + \partial_-^2h_{+i} + \partial_-\sigma\partial_-h_{+i}) + O(r)\\
R_{+i} &= \frac{1}{2}(\partial_-\sigma \partial_+h_{+i} + \partial_+\partial_-h_{+i}) + O(r)
\end{align}
As before, we use the components of the Ricci tensor to define fall-off constraints on the background metric $\met$, assuming that the matter energy-momentum tensor in the near-horizon region satisfies:
\begin{align}
T_{+-}  \,\,\,\hat{=}\,\,\, O(r^2)\label{eq-TAssumpGen1}\\
T_{-i}  \,\,\,\hat{=}\,\,\, O(r)\label{eq-TAssumpGen2}\\
T_{+i}  \,\,\,\hat{=}\,\,\, O(r)\label{eq-TAssumpGen3}\\
T_{ij}  \,\,\,\hat{=}\,\,\, O(r)\label{eq-TAssumpGen4}
\end{align}
This gives the following constraints on the metric:
\begin{align}
\partial_+\partial_-\sigma \,\,\,\hat{=}\,\,\, O(r)\label{eq-GenConst1}\\
\partial_-^2\sigma h_{+i} + \partial_-^2 h_{+i} + \partial_-\sigma\partial_- h_{+i} \,\,\,\hat{=}\,\,\, O(r)\label{eq-GenConst2}\\
\partial_-\sigma \partial_+ h_{+i} + \partial_+\partial_- h_{+i} \,\,\,\hat{=}\,\,\, O(r)\label{eq-GenConst3}\\
R - \frac{2n\Lambda}{n-2} \,\,\,\hat{=}\,\,\, O(r)\label{eq-GenConst4}
\end{align}
These conditions are necessary (but not sufficient) for $\met$ to satisfy the Einstein equations as $r\to 0$.

We now evaluate the Einstein-Hilbert action for the new metric $\met'$ and impose the fall-off constraints on the background metric $\met$. We work to leading order in $(r,\phi)$, where $\phi$ is defined as
\begin{equation}
\phi = \partial_+\xi^+ + \partial_-\xi^- + \xi^+\partial_+\sigma + \xi^-\partial_-\sigma
\end{equation}
as before. Although $\sigma$ (and therefore $\phi$) is now a function of $x^i$, the requirement that $\partial_i\partial_+\sigma=O(r)$ and $\partial_i\partial_-\sigma=O(r)$ means that to leading order $\phi$ is independent of $x^i$ and may be considered as a field on the $(\p,\m)$ submanifold. We find that once again all the dynamics in the $x^i$ coordinates disappear to $O(r^2)$, so we can integrate over these coordinates. We eventually obtain an action of the same form as $I_{\mathrm{hor}}$ given in (\ref{eq-action}) for the field $\phi$, but with a slightly different definition for the parameter $\lambda$ (see Appendix \ref{app-3} for details.)

As before, our result clearly applies in the case where there is no matter in the near-horizon region. And by the same reasoning as in Sec. \ref{sec-Action}, assumptions (\ref{eq-TAssumpGen2})-(\ref{eq-TAssumpGen3}) are likely to hold in very general cases, while assumptions (\ref{eq-TAssumpGen1}) and (\ref{eq-TAssumpGen4}) are more restrictive. There are likely to be background field configurations of physical interest where we have $T_{+-} \,\hat{=}\, O(r)$ and $T_{ij} \,\hat{=}\, O(1)$. But once again, even in this more general case, we will still preserve the constraints (\ref{eq-GenConst1})-(\ref{eq-GenConst3}) on the metric, and the only constraint that changes is (\ref{eq-GenConst4}), which becomes:
\begin{equation}
R - \frac{2n\Lambda-16\pi G T^\mu_\mu}{n-2} \,\,\,\hat{=}\,\,\, O(r)
\end{equation}
except in the case $n=2$, when we have $R\,\hat{=}\, O(r)$ as before. Working through the derivation of the dynamical action, we find that only the constraints (\ref{eq-GenConst1})-(\ref{eq-GenConst3}) are required until the very end of the computation, and thus we obtain the same result as in Sec. \ref{sec-Action}: the original Liouville-like action is modified by a potential term for $\phi$ proportional to $T^\mu_\mu$, and we have to know the precise form of $T_{\mu\nu}$ before evaluating this term. Once again, the dynamics of the field $\phi$ are unchanged by the addition of matter, and $\phi$ still becomes a free two-dimensional conformal field in an infinitesimal neighborhood of the horizon.

\section{Conclusion}\label{sec-Conclusion}

We have considered a very large class of horizons, including cosmological and acceleration horizons, and have explicitly derived the diffeomorphisms that preserve the characteristics of a horizon in a neighbourhood of the horizon. We then determined the dynamical action of these diffeomorphisms, and found that these degrees of freedom can be described by an effective two-dimensional Liouville-like theory that becomes conformal infinitesimally close to the horizon. 

We can now use this theory to study the thermodynamic properties of horizons. The form of $I_{\mathrm{hor}}$ is very suggestive: the central charge of the theory will be proportional to $\ar$, indicating that applying the Cardy formula may yield the Bekenstein-Hawking entropy. Now that we have an explicit action for the horizon degrees of freedom, it may also be possible to couple these degrees of freedom to scalar field matter and reproduce Hawking radiation.


\begin{appendix}

\section{The Ricci tensor}\label{app-1}

In this section we compute the components of the Ricci tensor for a general metric of the form (\ref{eq-conf}). We find:
\begin{align}
R_{+-} &= \partial_-\Gamma_{++}^+ - \partial_-\Gamma_{i+}^i - \Gamma_{-j}^i\Gamma_{i+}^j\nonumber\\
&= -g^{+-}\partial_+\partial_-\g + (g^{+-})^2\partial_-\g\partial_+\g\nonumber\\
&\,\,\,\,\,\,\,-\frac{1}{2}\left[ \partial_-(g^{ik}\partial_+g_{ik}) + \frac{1}{2}g^{im}g^{jk}\partial_-g_{jm}\partial_+g_{ik}\right ]\nonumber\\
&= -g^{+-}\partial_+\partial_-\g + (g^{+-})^2\partial_-\g\partial_+\g + O(r)\nonumber\\
&= -\partial_+\partial_-\sigma + O(r)\\
R_{++} &= -\partial_+\Gamma_{i+}^i + \Gamma_{i+}^i\Gamma_{++}^+ - \Gamma_{+i}^j\Gamma_{j+}^i\nonumber\\
&= -g^{ik}\partial_+^2 g_{ik} + g^{+-}g^{ik}\partial_+\g\partial_+ g_{ik} + O(r^2)\nonumber\\
&= -g^{ik}(\partial_+^2 g_{ik} + \partial_+\sigma\partial_+ g_{ik}) + O(r^2)\\
R_{--} &= -\partial_-\Gamma_{i-}^i + \Gamma_{i-}^i\Gamma_{--}^- - \Gamma_{-i}^j\Gamma_{j-}^i\nonumber\\
&= -g^{ik}\partial_-^2 g_{ik} + g^{+-}g^{ik}\partial_-\g\partial_- g_{ik} + O(r^2)\nonumber\\
&= -g^{ik}(\partial_-^2 g_{ik} + \partial_-\sigma\partial_- g_{ik})+O(r^2)\\
R_{ij} &= \tilde{R}_{ij} - g^{+-}\partial_+\partial_-g_{ij}+ \Gamma_{k+}^k\Gamma_{ij}^+ +  \Gamma_{k-}^k\Gamma_{ij}^- \nonumber\\
&\,\,\,\,\,\,\,- 2( \Gamma_{jk}^+\Gamma_{+i}^k + \Gamma_{j+}^k\Gamma_{ki}^+)\nonumber\\
&= \tilde{R}_{ij} - g^{+-}\partial_+\partial_-g_{ij} + O(r),
\end{align}
These equations lead to the constraints (\ref{eq-constraints1})-(\ref{eq-constraints4}). We can use (\ref{eq-constraints1})-(\ref{eq-constraints3}) to obtain the following useful expression:
\begin{align}\label{eq-Simple}
{g'}^{ij}\partial_+\partial_-g'_{ij} = \g' g^{+-}{g}^{ij}\partial_+\partial_-g_{ij} + O(r^2)
\end{align}
To see this, we calculate:
\begin{align}
&{g'}^{ij}\partial_-\partial_+g'_{ij} - \g'g^{+-}g^{ij}\partial_-\partial_+g_{ij}\nonumber\\
&\,\,\,\,\,\,\,\,\,\,= g^{ij}\left [\partial_-\partial_+ g'_{ij} - \g'g^{+-}\partial_-\partial_+g_{ij} \right ]\nonumber\\
&\,\,\,\,\,\,\,\,\,\,= g^{ij}\left [\partial_-\partial_+ g_{ij} - (1+\phi)\partial_-\partial_+g_{ij} +\partial_-\partial_+(\delta g_{ij})\right ]\nonumber\\
&\,\,\,\,\,\,\,\,\,\,= g^{ij}\left [-(\partial_+\xi^+ + \partial_-\xi^- + \xi^+\partial_+\sigma + \xi^-\partial_-\sigma)\partial_-\partial_+ g_{ij}\right ] \nonumber\\
&\hspace{1.5cm}+g^{ij}\partial_-\partial_+(\delta g_{ij})\nonumber\\
&\,\,\,\,\,\,\,\,\,\,= g^{ij}\left [\xi^+\partial_-\partial_+^2g_{ij} + \xi^-\partial_+\partial_-^2g_{ij}\right ]\nonumber\\
&\hspace{1.5cm}- g^{ij}(\xi^+\partial_+\sigma + \xi^-\partial_-\sigma)\partial_-\partial_+g_{ij}\nonumber\\
&\,\,\,\,\,\,\,\,\,\,= g^{ij}\xi^+(\partial_-\partial_+^2 g_{ij} - \partial_+\sigma \partial_+\partial_-g_{ij})\nonumber\\
&\hspace{1.5cm}+ g^{ij}\xi^-(\partial_+\partial_-^2 g_{ij} - \partial_-\sigma \partial_+\partial_-g_{ij})\label{eq-Simple1})
\end{align}
Using (\ref{eq-constraints2})-(\ref{eq-constraints3}), we find that
\begin{align}
\partial_-(\partial_+^2 g_{ik} + \partial_+\sigma\partial_+ g_{ik}) &= O(r^2)\\
\partial_+(\partial_-^2 g_{ik} + \partial_-\sigma\partial_- g_{ik}) &= O(r^2)
\end{align}
As we already have $\partial_+\partial_-\sigma = O(r)$ from (\ref{eq-constraints1}), this implies that
\begin{align}
\partial_-\partial_+^2 g_{ik} + \partial_+\sigma\partial_+\partial_- g_{ik} &= O(r^2)\\
\partial_+\partial_-^2 g_{ik} + \partial_-\sigma\partial_+\partial_- g_{ik} &= O(r^2)
\end{align}
Thus we find that (\ref{eq-Simple1}) is $O(r^2)$, which gives (\ref{eq-Simple}).

\section{The Dynamical Action}\label{app-2}

In this section we present the details of the calculation for deriving the dynamical action in Sec. \ref{sec-Action}. We begin with the expression for $R'_{+-}$:
\begin{align}
R'_{+-} &= -{g'}^{+-}\partial_+\partial_-\g' + ({g'}^{+-})^2\partial_-\g'\partial_+\g'\nonumber\\
&\,\,\,\,\,\,\,\,\,-\frac{1}{2}{g'}^{ij}\partial_-\partial_+ g'_{ij} + O(r^2)
\end{align}
We can simplify the last term using (\ref{eq-Simple}):
\begin{align}
R'_{+-} &= -{g'}^{+-}\partial_+\partial_-\g' + ({g'}^{+-})^2\partial_-\g'\partial_+\g'\nonumber\\
&\,\,\,\,\,\,\,\,\,-\frac{1}{2}\g' g^{+-}{g}^{ij}\partial_-\partial_+ g_{ij} + O(r^2)\label{eq-Ricci2}
\end{align}
The first two terms of $R'_{+-}$ can be evaluated in terms of $\phi$ and the background metric $\met$:
\begin{align}
&-{g'}^{+-}\partial_+\partial_-\g' + ({g'}^{+-})^2\partial_-\g'\partial_+\g'\nonumber\\
&\hspace{1cm}= \partial_+\phi\partial_-\phi\nonumber\\
&\hspace{1.4cm}+ (1+\partial_-\xi^- + \partial_+\xi^+)\partial_+\partial_-\sigma\nonumber\\
&\hspace{1.4cm}+ \xi^+\partial_+^2\partial_-\sigma + \xi^-\partial_+\partial_-^2\sigma\nonumber\\
&\hspace{1.4cm}+ O(\phi^3, r\phi^2)\label{eq-RicciNewApp}
\end{align}
We are ignoring terms of $O(\phi^3)$ and $O(r\phi^2)$ as we are working to leading order in $(r,\phi)$ and the leading terms in the action will be at most $O(\phi^2)$ or $O(r\phi)$.

We now evaluate $R'_{ij}$:
\begin{align}
R_{ij}' &= \tilde{R}'_{ij} - {g'}^{+-}\partial_+\partial_-g'_{ij} + O(r)
\end{align}
The boundary conditions on $\met$ give
\begin{align}
\tilde{R}'_{ij} &= \tilde{R}_{ij} + O(r)
\end{align}
and therefore 
\begin{align}
{g'}^{ij}R'_{ij} &= g^{ij}\tilde{R}_{ij} - {g'}^{ij}{g'}^{+-}\partial_+\partial_-g'_{ij}+ O(r)\label{eq-RijNewApp}
\end{align}
We can now evaluate $R'$ using (\ref{eq-RicciNewApp}) and (\ref{eq-RijNewApp}):
\begin{align}
R' &= 2{g'}^{+-}R'_{+-} + {g'}^{ij}R'_{ij}\nonumber\\
&= 2{g'}^{+-}\left [\partial_+\phi\partial_-\phi-\partial_+\partial_-\sigma(1+\partial_+\xi^+ + \partial_-\xi^-)\right ]\nonumber\\
&\hspace{0.4cm}-2{g'}^{+-}(\xi^+\partial_+^2\partial_-\sigma + \xi^-\partial_+\partial_-^2\sigma)\nonumber\\
&\hspace{0.4cm}-2{g}^{+-}{g}^{ij}\partial_-\partial_+g_{ij} + {g}^{ij}\tilde{R}^{ij} + O(r)\label{eq-RNew}
\end{align}
We can simplify this expression further by rewriting the constraint (\ref{eq-constraints4}) as:
\begin{align}
R &= -2g^{+-}\partial_+\partial_-\sigma -2{g}^{+-}{g}^{ij}\partial_-\partial_+g_{ij} + {g}^{ij}\tilde{R}^{ij}\nonumber\\
&\,\,\hat{=}\,\,\, \frac{2n\Lambda}{n-2} + O(r)
\end{align}
Using the above expression to simplify (\ref{eq-RNew}) allows us to write $R' - 2\Lambda$ as:
\begin{align}
R' - 2\Lambda &= 2{g'}^{+-}\partial_+\phi\partial_-\phi\label{eq-Kin}\\
&\hspace{0.4cm}-2{g'}^{+-}\partial_+\partial_-\sigma(1+\partial_+\xi^+ + \partial_-\xi^-)\label{eq-R1}\\
&\hspace{0.4cm}-2{g'}^{+-}(\xi^+\partial_+^2\partial_-\sigma  + \xi^-\partial_+\partial_-^2\sigma)\label{eq-R2}\\
&\hspace{0.4cm}+ 2g^{+-}\partial_+\partial_-\sigma\label{eq-R3}\\
&\hspace{0.4cm}+\frac{4\Lambda}{n-2} + O(r)\label{eq-Const}
\end{align}
Note that this is the only step where the constraint (\ref{eq-constraints4}) was required: for all the other computations, the constraints (\ref{eq-constraints1})-(\ref{eq-constraints3}) were sufficient. 

In order to obtain the Einstein-Hilbert action, we multiply $(R' - 2\Lambda)$ by the metric determinant $\sqrt{-g'}$, which to leading order is:
\begin{align}
\sqrt{-g'} &= \sqrt{\tilde{g}}\sqrt{-\hat{g}}(1+\phi)
\end{align}
Multiplying this with the first term (\ref{eq-Kin}) of $(R' - 2\Lambda)$ then gives:
\begin{align}
\sqrt{\tilde{g}}\sqrt{-\hat{g}} \left (2g^{+-}\partial_+\phi\partial_-\phi\right ) &= \sqrt{\tilde{g}}\sqrt{-\hat{g}} \left (g^{ab}\partial_a\phi\partial_b\phi\right )\nonumber\\
&= \sqrt{\tilde{g}}\sqrt{-\hat{g}} \left (\partial_a\phi\partial^a\phi\right )\label{eq-Final1}
\end{align}
where the index $a \in (\p,\m)$ and $\hat{g}_{ab}$ is the induced metric on the $(\p,\m)$ submanifold. 

Similarly, multiplying $\sqrt{-g'}$ with the terms (\ref{eq-R1})-(\ref{eq-R3}) gives:
\begin{align}
&-2\sqrt{\tilde{g}}\left (\xi^+\partial_+^2\partial_-\sigma  + \xi^-\partial_+\partial_-^2\sigma \right )\nonumber\\
&\hspace{0.5cm}-2\sqrt{\tilde{g}} \partial_+\partial_-\sigma\left ( 1+\partial_+\xi^+ + \partial_-\xi^-\right ) \nonumber\\
&\hspace{0.5cm}+ 2\sqrt{\tilde{g}}\partial_+\partial_-\sigma(1+\phi)
\end{align}
This expression simplifies to
\begin{align}
&2\sqrt{\tilde{g}}\left [\partial_+\partial_-\sigma (\xi^+\partial_+\sigma + \xi^-\partial_-\sigma) -\xi^+\partial_+^2\partial_-\sigma  - \xi^-\partial_+\partial_-^2\sigma\right ]\label{eq-Final2a}
\end{align}
We now use integration by parts to rewrite (\ref{eq-Final2a}) as
\begin{align}
&= 2\sqrt{\tilde{g}}\partial_+\partial_-\sigma (\xi^+\partial_+\sigma + \xi^-\partial_-\sigma + \partial_+\xi^+ + \partial_-\xi^-)\nonumber\\
&= -\sqrt{\tilde{g}}\left (\g \phi\hat{R}\right)\nonumber\\
&= -\sqrt{\tilde{g}}\sqrt{\hat{g}}\phi\hat{R}\label{eq-Final2b},
\end{align}
where $\hat{R}$ is the Ricci scalar corresponding to the induced metric $\hat{g}_{ab}$ on the $(\p,\m)$ submanifold, and $\hat{R}$ is given by
\begin{equation}
\hat{R} = -2g^{+-}\partial_+\partial_-\sigma\label{eq-hatR}
\end{equation}

Finally, multiplying $\sqrt{-g'}$ with the last term (\ref{eq-Const}) of $(R'-2\Lambda)$ gives:
\begin{align}
\sqrt{\tilde{g}}\sqrt{-\hat{g}}\frac{4\Lambda}{n-2}(1+\phi)\label{eq-Final3}
\end{align}
Combining (\ref{eq-Final1}), (\ref{eq-Final2b}), and (\ref{eq-Final3}) gives:
\begin{align}
\sqrt{-g'}(R' - 2\Lambda) &= \sqrt{\tilde{g}}\sqrt{-\hat{g}}\left [\partial_a\phi\partial^a\phi - \phi\hat{R} + \lambda(1+\phi)\right ]\nonumber
\end{align}
with $\lambda = \frac{4\Lambda}{n-2}$. The only dependence on the $x^i$ coordinates is in the metric determinant $\sqrt{\tilde{g}}$. Thus we may integrate over these coordinates in the Einstein-Hilbert action, to obtain the final dynamical action:
\begin{align}
I_{\mathrm{hor}} = \frac{\ar}{16\pi G} \int\rd{^2x} \sqrt{-\hat{g}}\,\, \left (\partial_a\phi\partial^a\phi -\phi\hat{R}+ \lambda (1+\phi)\right )\nonumber
\end{align}
where $\ar$ is the cross-sectional area of the horizon, and the variables of integration are $(\p,\m)$.

\section{More General Horizons}\label{app-3}

In this section we present the details of the calculations used to derive the dynamical action in Sec. \ref{sec-General}, where we consider a much larger class of horizons that can be written in the form (\ref{eq-conf2}). As in Sec. \ref{sec-Action}, we work to leading order in $(r,\phi)$.

Evaluating the Ricci tensor for metrics of the form (\ref{eq-conf2}), we find:
\begin{align}
R_{+-} &= -\partial_-\Gamma_{++}^+ - \partial_-\Gamma_{i+}^i +\partial_i\Gamma_{+-}^i + \Gamma_{-j}^i\Gamma_{i+}^j + O(r^2)\nonumber\\
&= -\partial_+\partial_-\sigma + O(r)\\
R_{-i} &= \partial_-\Gamma_{-i}^- - \partial_i\Gamma_{--}^- + O(r)\nonumber\\
&= \frac{1}{2}(\partial_-^2\sigma h_{+i} + \partial_-^2h_{+i} + \partial_-\sigma\partial_-h_{+i}) + O(r)\\
R_{+i} &= \partial_+\Gamma_{+i}^+ - \partial_i\Gamma_{++}^+ + O(r)\nonumber\\
&= \frac{1}{2}(\partial_-\sigma \partial_+h_{+i} + \partial_+\partial_-h_{+i}) + O(r)
\end{align}
Thus, in order for $\met$ to be able to satisfy the Einstein equations as $r\to 0$, we must have
\begin{align}
\partial_+\partial_-\sigma \,\,\,\hat{=}\,\,\, O(r)\\
\partial_-^2\sigma h_{+i} + \partial_-^2 h_{+i} + \partial_-\sigma\partial_- h_{+i} \,\,\,\hat{=}\,\,\, O(r)\\
\partial_-\sigma \partial_+ h_{+i} + \partial_+\partial_- h_{+i} \,\,\,\hat{=}\,\,\, O(r)\\
R - \frac{2n\Lambda}{n-2} \,\,\,\hat{=}\,\,\, O(r)
\end{align}
giving constraints on $\met$.

In order to evaluate the Einstein-Hilbert action for $\met'$, we need to compute
\begin{align}
R' &= 2(g^{'+-}R'_{+-} + g^{'i-}R'_{i-}) + g'^{ij}R'_{ij}
\end{align}
to leading order in $(r,\phi)$.
We first find
\begin{align}
R_{+-} &= -\partial_-\Gamma_{++}^+\nonumber\\
&\hspace{0.5cm}+ \frac{1}{2} \biggl [g^{ij}(\partial_-\partial_j g_{+i} - \partial_i\partial_j\g - \partial_-\partial_+g_{ij})\nonumber\\
&\hspace{0.7cm} + (\partial_-g_{+j} - \partial_j\g)g^{ij}(\tilde{\Gamma}^k_{ki} + \frac{1}{2}g^{k-}\partial_-g_{ik})\nonumber\\
&\hspace{0.7cm} + (\partial_-g_{+j} - \partial_j\g)(\partial_-g^{j-} + \partial_i g^{ij})\nonumber\\
&\hspace{0.7cm} - g^{i-}(\partial_i\partial_-\g - \partial_-^2 g_{+i}) \biggr ] + O(r^2)\\
R_{-i} &= \frac{1}{2}g^{+-}(\partial_-^2 g_{+i} - \partial_-\partial_i \g)\nonumber\\
&\hspace{0.7cm} + \frac{1}{2}(g^{+-})^2(\partial_-\g)(\partial_i\g - \partial_- g_{+i})\nonumber\\
&\hspace{0.7cm} + O(r)
\end{align}
where as before, $\tilde{A}$ denotes a quantity $A$ computed with respect to the induced metric $\tilde{g}_{ij}$ on the $r=0$ hypersurface. This gives:
\begin{align}
&2(g^{i-}R_{i-} + g^{+-}R_{+-})\nonumber\\
&\hspace{0.2cm} = g^{i-}g^{+-}(\partial_-^2g_{+i} - \partial_i\partial_-\g)\nonumber\\
&\hspace{0.5cm} + g^{i-}(g^{+-})^2(\partial_-\g\partial_i\g - \partial_-\g\partial_i g_{+i})\nonumber\\
&\hspace{0.5cm} -2\partial_-\Gamma_{++}^+\nonumber\\
&\hspace{0.5cm} + g^{ij}g^{+-}(\partial_-\partial_j g_{+i} - \partial_i \partial_j \g - \partial_- \partial_+ g_{ij})\nonumber\\
&\hspace{0.5cm} + g^{+-}(\partial_- g_{+j} - \partial_-\g)g^{ij}(\tilde{\Gamma}^k_{ki} + \frac{1}{2}g^{k-}\partial_- g_{ik})\nonumber\\
&\hspace{0.5cm} + g^{+-}(\partial_-g_{+j} - \partial_-\g)(\partial_i g^{ij} + \partial_- g^{j-})\nonumber\\
&\hspace{0.5cm} + O(r)\label{eq-Gen1}\end{align}
Similarly, we can compute:
\begin{align}
g^{ij}R_{ij} &= -g^{ij}g^{+-}(\partial_+\partial_-g_{ij} + \partial_i\partial_j\g)\nonumber\\
&\hspace{0.5cm} + \frac{1}{2}g^{ij}g^{+-}(\partial_-\partial_ig_{+j} +\partial_-\partial_j g_{+i})\nonumber\\
&\hspace{0.5cm} + \frac{1}{2}g^{ij}(\partial_j\g\partial_i\g - \partial_-g_{+j}\partial_-g_{+i})\nonumber\\
&\hspace{0.5cm} + g^{ij}g^{+-}\partial_k\g\tilde{\Gamma}_{ij}^k + g^{ij}\tilde{R}_{ij} + O(r)\label{eq-Gen2}
\end{align}
We can now compute $\g'R'$ by varying all the quantities in (\ref{eq-Gen1}) and (\ref{eq-Gen2}) under a diffeomorphism $\xi$ of the form (\ref{eq-diff}). We write:
\begin{align}
\g'R' &= -2\g'\partial_-\Gamma_{++}^{'+} + B(\phi, \met)
\end{align}
for some function $B(\phi, \met)$. If we can show that
\begin{align}\label{eq-Key}
B(\phi, \met) - \g'(R + 2g^{+-}\partial_+\partial_-\sigma) = O(r^2),
\end{align}
then we can write:
\begin{align}
\g'R' &= -2\g'\partial_-\Gamma_{++}^{'+} + \g' (R + 2g^{+-}\partial_+\partial_-\sigma) + O(r^2)\nonumber\\
&= 2\g'(-\partial_-\Gamma_{++}^{'+} + g^{+-}\partial_+\partial_-\sigma)\nonumber\\
&\hspace{0.5cm} + \g'\frac{2n\Lambda}{n-2} + O(r^2)\label{eq-GenFinalR}
\end{align}
Calculating the first term in the above expression, we find:
\begin{align}
&2\g'(-\partial_-\Gamma_{++}^{'+} + g^{+-}\partial_+\partial_-\sigma)\nonumber\\
&\hspace{0.5cm} = 2\g'(-{g'}^{+-}\partial_+\partial_-\g' + ({g'}^{+-})^2\partial_-\g'\partial_+\g')\nonumber\\ 
&\hspace{1cm} + 2\g'g^{+-}\partial_+\partial_-\sigma
\end{align}
From our work in Appendix \ref{app-1} we can see that this is equivalent to
\begin{align}\label{eq-GenFinal1}
\sqrt{-\hat{g}}(\partial_a\phi\partial^a \phi - \phi\hat{R})
\end{align}
Looking at the form of the metric in (\ref{eq-conf2}), we might be concerned that $x^i$-dependence will enter through $\hat{g}$ and $\hat{R}$, when we are trying to define an action on the $(x_+, x_-)$ submanifold. However, recall that $g_{+-} = e^{\sigma(u,r,x^i)}$ with $\sigma = \ln r + \sigma_0(u) + \sigma_1(x^i) + O(r)$. Thus the $x^i$-dependence in $\sqrt{-\hat{g}} = g_{+-}$ cancels with the $x^i$ dependence in the kinetic term $\partial_a\phi\partial^a\phi = 2g^{+-}\partial_+\phi\partial_-\phi$. We have already established that $\phi$ can be interpreted as a field on the $(x_+, x_-)$ submanifold to leading order, so the kinetic term can be defined on the $(x_+, x_-)$ submanifold. A similar cancellation of the $x^i$-dependence occurs for the $\hat{R}$ term, as can be seen from the definition (\ref{eq-hatR}) of $\hat{R}$. As a result, we can interpret the expression (\ref{eq-GenFinal1}) as a quantity defined on the $(x_+, x_-)$ submanifold, by redefining:
\begin{align}
\hat{g}_{+-} := e^{\ln r + \sigma_0(u) + O(r)}
\end{align}
and
\begin{align}
\hat{R} = -2\hat{g}^{+-}\partial_+\partial_-\sigma
\end{align}
Substituting (\ref{eq-GenFinal1}) back into (\ref{eq-GenFinalR}) we find:
\begin{align}
\g'(R' - 2\Lambda) &= \sqrt{-\hat{g}}(\partial_a\phi\partial^a \phi - \phi\hat{R} + \tilde{\lambda}(1+\phi))
\end{align}
where the $x^i$-dependence of $\sqrt{-\hat{g}}$ only appears in the last term, and has been incorporated into $\tilde{\lambda} := \frac{4\Lambda e^{\sigma_1(x^i)}}{n-2}$. This gives:
\begin{align}
\sqrt{-g'}(R' - 2\Lambda) &= \sqrt{\tilde{g}}\sqrt{-\hat{g}}(\partial_a\phi\partial^a \phi - \phi\hat{R} + \tilde{\lambda}(1+\phi))
\end{align}
to leading order in $(r,\phi)$. We can integrate over the $x^i$ coordinates, and the first two terms (the kinetic term and the $\phi\hat{R}$ coupling term) give the same results as in Sec. \ref{sec-Action}. The last term is slightly different, as $\tilde{\lambda}$ must be included in the integral over the $x^i$ coordinates. In the end we obtain an action of the same form as $I_{\mathrm{hor}}$ in Sec. \ref{sec-Action}, but with a slightly different definition for the parameter $\lambda$:
\begin{align}
\lambda := \frac{1}{a_{\Delta}} \int \mathrm{d}^{n-2}x^i\,\,\, \tilde{\lambda}\sqrt{\tilde{g}},
\end{align}
where the integral is carried out over the $(n-2)$ coordinates $x^i$, and $a_{\Delta}$ is the cross-sectional area of the horizon.

Now all we have to do is derive (\ref{eq-Key}) in order to obtain our final result. By direct computation of $\g'R'$, we find:
\begin{align}
&B(\phi, \met) - \g'(R + 2g^{+-}\partial_+\partial_-\sigma)\nonumber\\
&\hspace{0.2cm}= -\phi g^{ij}\tilde{\Gamma}^k_{ij}\partial_k\g + g^{ij}\tilde{\Gamma}_{ij}^k\partial_k(\delta\g)\label{eq-Term3}\\
&\hspace{0.3cm}-\phi\partial_i g^{ij}(\partial_- (\delta g_{+j}) - \partial_j(\delta \g))\label{eq-Term9a}\\
&\hspace{0.3cm}+ \partial_i g^{ij}(\partial_i (\delta g_{+j}) - \partial_j(\delta \g))\label{eq-Term9b}\\
&\hspace{0.3cm}-\phi g^{ij}\tilde{\Gamma}^k_{ki}(\partial_- g_{+j} - \partial_j\g)\label{eq-Term4a}\\
&\hspace{0.3cm}+ g^{ij}\tilde{\Gamma}^k_{ki}(\partial_i (\delta g_{+j}) - \partial_j(\delta \g))\label{eq-Term4b}\\
&\hspace{0.3cm}-2\phi g^{ij}(\partial_-\partial_j g_{+i} - \partial_i\partial_j\g - \partial_-\partial_+ g_{ij})\label{eq-Term2a}\\
&\hspace{0.3cm}+ 2g^{ij}(\partial_-\partial_j (\delta g_{+i}) - \partial_i\partial_j(\delta \g) - \partial_-\partial_+ (\delta g_{ij}))\label{eq-Term2b}\\
&\hspace{0.3cm}+ \frac{1}{2}g^{ij}g^{+-}(1-\phi)(\partial_j \g'\partial_i\g' - \partial_- g_{+j}'\partial_- g_{+i}')\label{eq-Term5}\\
&\hspace{0.3cm}- \frac{1}{2}g^{ij}g^{+-}(1+\phi)(\partial_j \g\partial_i\g - \partial_- g_{+j}\partial_- g_{+i})\label{eq-Term5b}\\
&\hspace{0.3cm}+ 2g^{'i-}(\partial_-^2 g_{+i}' - \partial_-\partial_i \g')\label{eq-Term6}\\
&\hspace{0.3cm}- 2g^{i-}(1+\phi)(\partial_-^2 g_{+i} - \partial_-\partial_i \g)\label{eq-Term6b}\\
&\hspace{0.3cm}+ g^{'i-}g^{+-}(1-\phi)\partial_-\g'(\partial_i\g' - \partial_-g_{+i}')\label{eq-Term7}\\
&\hspace{0.3cm}- g^{i-}g^{+-}(1+\phi)\partial_-\g(\partial_i\g - \partial_-g_{+i})\label{eq-Term7b}\\
&\hspace{0.3cm}+ \partial_-g^{'j-}(\partial_-g_{+j}' - \partial_j\g')\label{eq-Term8}\\
&\hspace{0.3cm}- \partial_-g^{j-}(1+\phi)(\partial_-g_{+j} - \partial_j\g)\label{eq-Term8b}
\end{align}
We simplify the above expression by applying the constraints (\ref{eq-GenConst1})-(\ref{eq-GenConst3}) to the background metric $\met$. To simplify the terms (\ref{eq-Term3}):
\begin{align}
-\phi\partial_k\g + \partial_k(\delta\g) &= -\phi \partial_k\g + \partial_k(\phi\g)\nonumber\\
&= \g\partial_k\phi = O(r^2),\nonumber
\end{align}
as $\g = O(r)$ and $\partial_k\phi = O(r)$ due to the conditions $\partial_{\pm}\partial_i\sigma = O(r)$. It follows that the terms (\ref{eq-Term3}) combine to give a quantity that is $O(r^2)$.

In order to simplify the terms (\ref{eq-Term9a})-(\ref{eq-Term9b}) as well as (\ref{eq-Term4a})-(\ref{eq-Term4b}), consider:
\begin{align}
&-\phi(\partial_- (\delta g_{+j}) - \partial_j(\delta \g))+ (\partial_i (\delta g_{+j}) - \partial_j(\delta \g))\nonumber\\
&\hspace{0.2cm}= \xi^+\g(\partial_-\sigma\partial_+h_{+i} + \partial_-\partial_+ h_{+i})\nonumber\\
&\hspace{0.4cm}+ \xi^-\g(\partial_-^2\sigma h_{+i} + \partial_-\sigma\partial_- h_{+i} + \partial_-^2 h_{+i})\nonumber\\
&\hspace{0.2cm}= O(r^2)
\end{align}
by the constraints (\ref{eq-GenConst2})-(\ref{eq-GenConst3}) on $\met$. It follows that the terms (\ref{eq-Term9a})-(\ref{eq-Term4b}) combine to give a quantity that is $O(r^2)$. The terms (\ref{eq-Term2a})-(\ref{eq-Term2b}) simplify in the same way to give a quantity that is $O(r^2)$.

We are left with the terms (\ref{eq-Term5})-(\ref{eq-Term8b}). In order to simplify these terms, we use the fact that
\begin{align}
{g'}^{-i} &= -g^{+-}(1-\phi)g'_{+j}g^{ij} + O(r)\\
&= (1-\phi)g^{-i} - (1-\phi)g^{+-}g^{ij}\delta g_{+j} + O(r)
\end{align}
Direct computation and the application of the constraints (\ref{eq-GenConst2})-(\ref{eq-GenConst3}) shows that the terms (\ref{eq-Term5})-(\ref{eq-Term8b}) also combine to give a quantity that is $O(r^2)$. So finally, we find that (\ref{eq-Key}) holds. Note that constraint (\ref{eq-GenConst4}) was not required for these computations.
\end{appendix}


\begin{thebibliography}{21}
\bibitem{Bekenstein} J.~D.~Bekenstein, Phys. Rev. D7 (1973), 2333.
\bibitem{Hawking} S.~W.~Hawking, Nature 248 (1974), 30.
\bibitem{Unruh} W.~G.~Unruh, Phys. Rev. D14 (1976), 870, G.~W.~Gibbons and S.~W.~Hawking, Phys. Rev. D15 (1977), 2738.
\bibitem{CarlipReview} S.~Carlip, Lect. Notes Phys. 769:89-123 (2009), arXiv:0807.4520 [gr-qc].
\bibitem{StromingerVafa} A.~Strominger and C.~Vafa, Phys. Lett. B379 (1996), 99, arXiv:hep-th/9601029.
\bibitem{LQG} C.~Rovelli, ``Loop Quantum Gravity,'' \textit{Living Rev. Relativity}, 11 (2008), 5. URL (cited on March 17th, 2011): http://www.livingreviews.org/lrr-2008-5.
\bibitem{StromingerCarlip} A.~Strominger, JHEP 9802 (1998), 009, arXiv:hep-th/9712251, S.~Carlip, Phys. Rev. Lett. 82 (1999), 2828, arXiv:hep-th/9812013.
\bibitem{BrownHenneaux} J.~D.~Brown and M.~Henneaux, Commun. Math. Phys. 104 (1986), 207.
\bibitem{Dreyer} O.~Dreyer, A.~Ghosh, and J.~Wi\'{s}niewski, Class. Quant. Grav. 18 (2001), 1929, arXiv:hep-th/0101117.
\bibitem{Silva} S.~Silva, Class. Quant. Grav. 19(15):3947-3961, (2002), arXiv:hep-th/0204179.
\bibitem{Koga} J-I.~Koga, Phys. Rev. D64 (2001), 124012, arXiv:gr-qc/0107096.
\bibitem{Kang} G.~Kang, J-I.~Koga, and M-I.~Park. Phys. Rev. D70 (2004), 024005. arXiv:hep-th/0402113.
\bibitem{CarlipAdS} S.~Carlip, Class. Quant. Grav. 22 (2005), 3055, arXiv:gr-qc/0501033.
\bibitem{Aros} R.~Aros, M.~Romo, and N.~Zamorano, Phys. Rev. D75 (2007), 067501, arXiv:hep-th/0612028.
\bibitem{Carlip21} S.~Carlip, Phys. Rev. D51 (1995), 632, arXiv:gr-qc/9409052.
\bibitem{Padmanabhan} T.~Padmanabhan, Rep. Prog. Phys. 73 (2010), 046901, [arXiv:0911.5004].
\bibitem{Solodukhin} S.~N.~Solodukhin, Phys. Lett. B454, 213 (1999), arXiv:hep-th/9812056.
\bibitem{DiasLemos} G.~A.~S.~Dias and J.~P.~S.~Lemos, Phys. Rev. D74 (2006), 044024, arXiv:hep-th/0602144.
\bibitem{Giacomini} A.~Giacomini and N.~Pinamonti, JHEP 0302 (2003), 014, arXiv:gr-qc/0301038.
\bibitem{Rodriguez} L.~Rodriguez and T.~Yildirim, Class. Quant. Grav. 27 (2010), 155003, arXiv:1003.0026 [hep-th].
\bibitem{Ashtekar} A.~Ashtekar and B.~Krishnan, ``Isolated and Dynamical Horizons and Their Applications,'' \textit{Living Rev. Relativity}, 7 (2004), 10. URL (cited on March 17th, 2011): http://www.livingreviews.org/lrr-2004-10.
\bibitem{Booth} I.~Booth, Can. J. Phys. 83 (2005), 1073, arXiv:gr-qc/0508107.
\bibitem{Wald2} H.~Friedrich, I.~Racz, and R.~M.~Wald, Commun. Math. Phys. 204 (2001), 691, arXiv:gr-qc/9811021.
\bibitem{FeffermanGraham} C.~Fefferman and C.~R.~Graham, In: \textit{Elie Cartan et les Mathematiques d'aujour'hui}, Asterisque, hors serie (1985) 95.
\bibitem{Medved1} A.~J.~M.~Medved, D.~Martin, and M.~Visser, Class. Quant. Grav. 21 (2004), 3111, arXiv:gr-qc/0402069.
\bibitem{Medved2} A.~J.~M.~Medved, D.~Martin, and M.~Visser, Phys. Rev. D70 (2004), 024009, arXiv:gr-qc/0403026.
\bibitem{Wald1} R.~M.~Wald, {\sl General Relativity}, The University of Chicago Press, (1984).
\bibitem{BoothVar} I.~S.~Booth, Class. Quant. Grav. 18 (2001), 4239, arXiv:gr-qc/0105009.
\bibitem{AshtekarVar} A.~Ashtekar, S.~Fairhurst, and B.~Krishnan, Phys. Rev. D62 (2000), 104025, arXiv:gr-qc/0005083.
\bibitem{Seiberg} N.~Seiberg, Prog. Theor. Phys. Suppl. 102, 319 (1990).
\end{thebibliography}
\end{document}